\documentclass{iau}
\usepackage{graphicx}
\newcommand{\lsi}{LS~I~$+$61$^{\circ}$303}

\title[Unveiling the super-orbital modulation of \lsi\ in X-rays] 
{Unveiling the super-orbital modulation of \lsi\ in X-rays}

\author[Jian Li \& Diego F. Torres]   
{Jian Li $^1,^2$
  Diego F. Torres$^2$, Shu Zhang$^1$, Daniela Hadasch$^2$, Nanda Rea$^2$, G. Andrea Caliandro$^2$, Yupeng Chen$^1$, \and Jianmin Wang$^1$}

\affiliation{$^1$ Key Laboratory for Particle Astrophysics, Institute of High Energy Physics, Chinese Academy of Sciences, 19B Yuquan Road, Beijing 100049, China,  email: {\tt jianli@ihep.ac.cn} \\[\affilskip]
$^2$Institut de Ci\`encies de l'Espai (IEEC-CSIC),
Campus UAB,  Torre C5, 2a planta, 08193 Barcelona, Spain\\[\affilskip]
}

\pubyear{2012}
\volume{290}  
\jname{Feeding compact objects: Accretion on all scales}
\editors{C.M. Zhang, T. Belloni, M. M\'endez \& S.N. Zhang, eds.}

\begin{document}

\maketitle

\begin{abstract}

We found evidence for the super--orbital modulation in the X-ray emission of \lsi\ from the longest monitoring date by the \emph{RXTE}. The time evolution of the modulated fraction in the orbital light curves can be well fitted with a sinusoidal function having a super-orbital period of 1667 days. However, we have found a 281.8$\pm$44.6 day shift between the super-orbital variability found at radio frequencies and our X-ray data. We also find a super-orbital modulation in the maximum count rate of the orbital light curves, compatible with the former results, including the shift.


\end{abstract}

\firstsection 

\section{Introduction}

\lsi\ is one of the elite $\gamma$--ray binaries. Its nature is still under debate,
 with rotationally powered pulsar-composed systems (see Maraschi $\&$ Treves
1981; Dubus 2006) and microquasar jets (see Bosch-Ramon
$\&$ Khangulyan 2009 for a review) being discussed. Long-term monitoring of
the source is a key ingredient to disentangle differences in behavior
which could point to the underlying source nature.
Here, we report on the analysis of \emph{RXTE/PCA} monitoring
observations of \lsi\ and the possible super-orbital modulation
of the X-ray emission.

\section{Observations and Results}

Our data set includes 473 \emph{RXTE/PCA} pointed observations from 2007 August 28 to 2011 September 15.
The analysis is performed using the standard \emph{RXTE/PCA} criteria.
Only PCU2 has been used for the analysis. Our
 count rate values are given for an energy range of 3-30 keV.
In order to remove the influence of several kilosecond-long flares, we cut all observations that presented a
larger count rate than three times the average.

\begin{figure*}[t]
\centering
  \includegraphics[angle=0, scale=0.334] {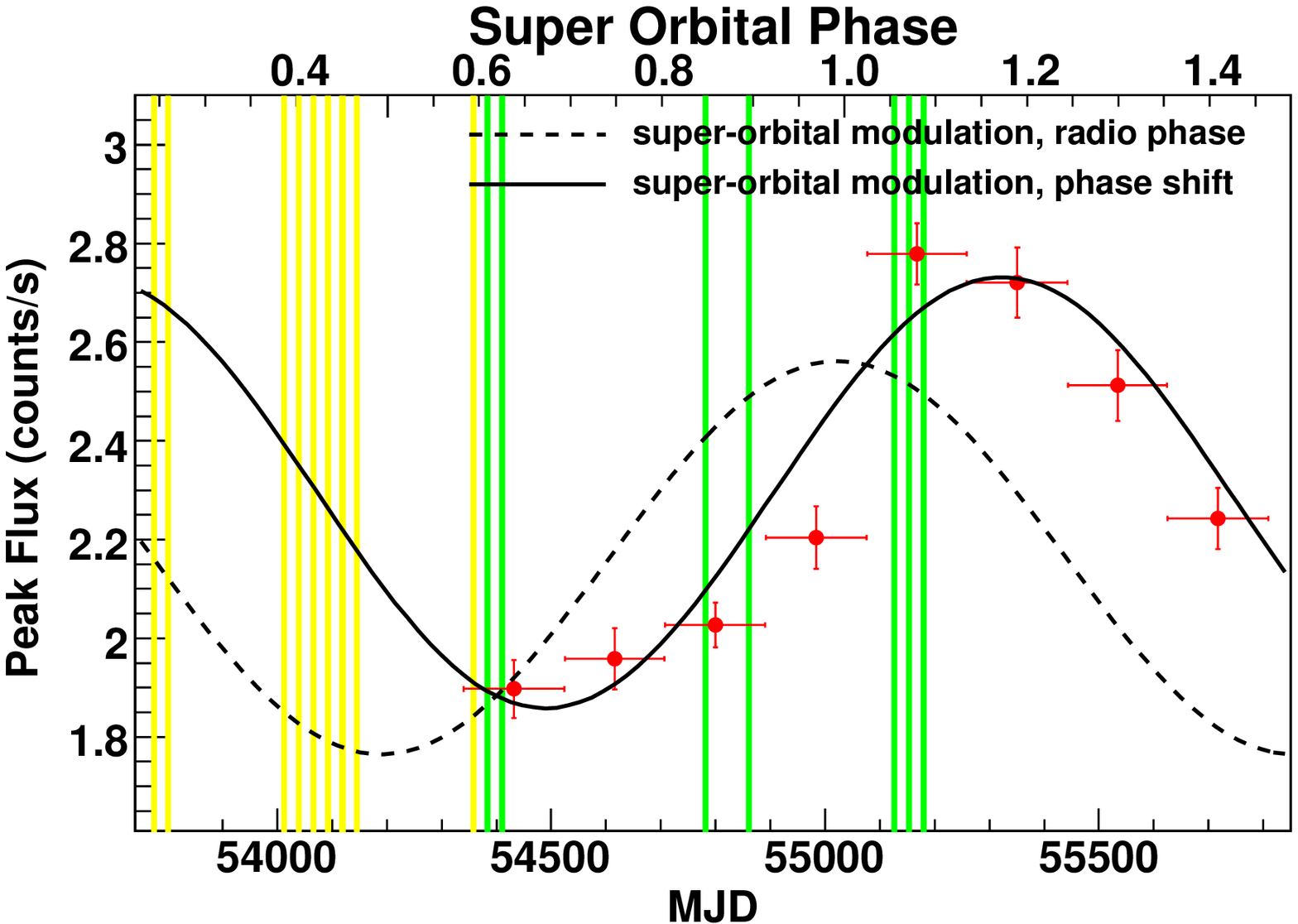}
  \includegraphics[angle=0, scale=0.334] {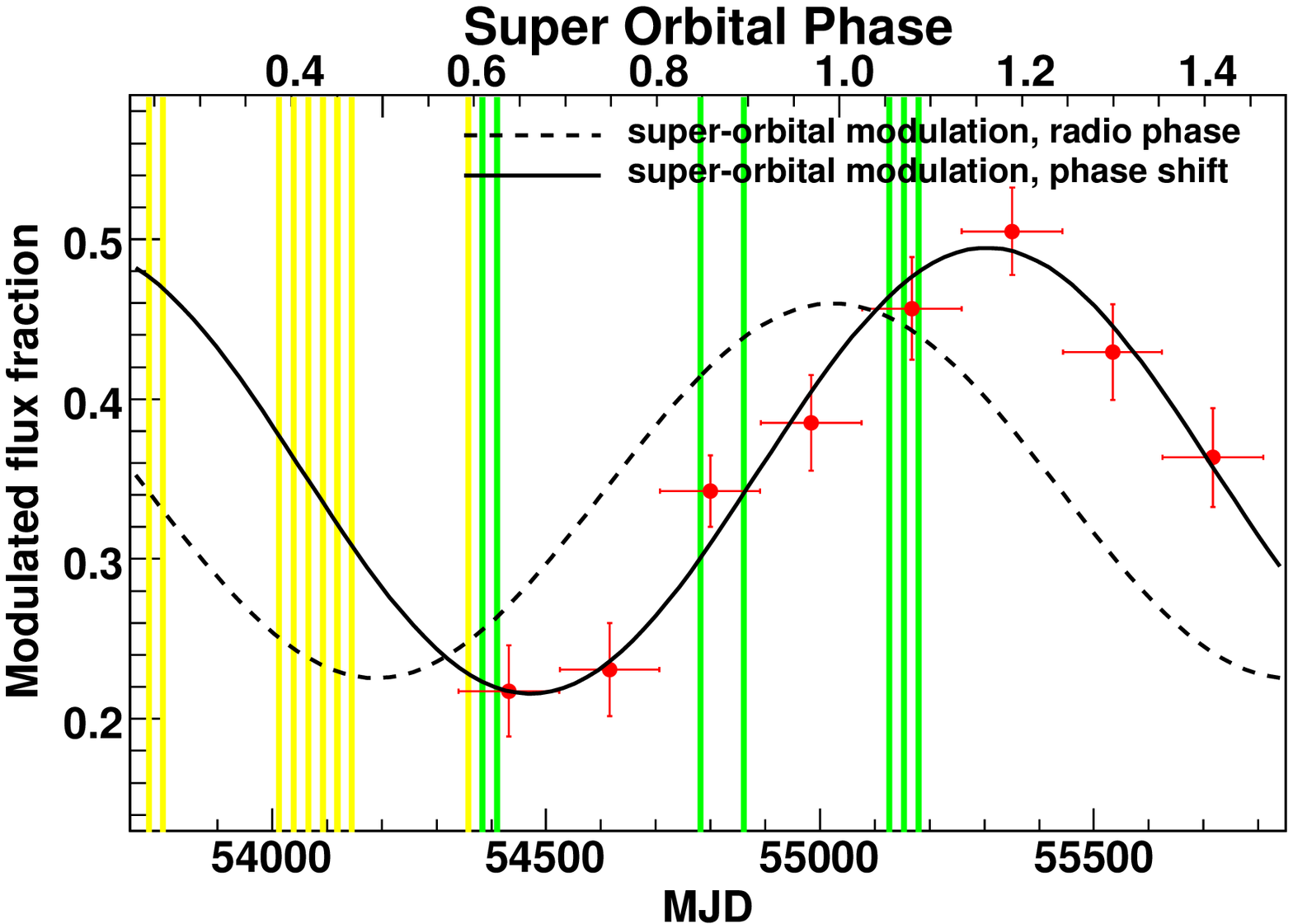}
 \caption{Left: Peak count rate of the X-ray emission from \lsi\ as a function of time and the super-orbital phase.
 Right: modulated fraction, see text for details. The dotted line shows the sine fitting to the modulated flux fraction and peak flux with a period and phase fixed at the radio parameters (from Gregory 2002).
The solid curve stands for sinusoidal fit obtained by fixing the period at the 1667 days value, but letting the phase vary. The time bin corresponds to six months. The colored boxes represent the times of the TeV observations that covered the broadly-defined apastron region. The boxes in green denote the times when TeV observations are in low state while boxes in yellow are TeV observations in high state.
}
\label{super-orb}
\end{figure*}

\begin{table}[t]
\scriptsize
\begin{center}
\label{tab1}
\caption{Reduced $\chi^2$ for fitting different models to the modulation fraction and the peak flux in X-rays.
}
\vspace{5pt}
\small
\begin{tabular}{lllll}
\hline
 & Constant & Linear & Radio & Shifted\\
\hline
Modulation Fraction & 88.2 / 7   & 38.0 / 6   &  42.1 / 6 & 1.1 / 5  \\
Peak Flux & 212.8 / 7   &  114.8 / 6  &  91.8 / 6 & 4.9 / 5  \\
\hline
\end{tabular}
\end{center}
\end{table}

Given a six-month time bin, we take the peak X-ray flux in orbital lightcurve and
compute the modulated flux fraction. The latter is defined as
$(c_{max} - c_{min})/(c_{max} + c_{min})$, where $c_{max}$ and $c_{min}$
are the maximum and minimum count rates in the 3-30 keV orbital lightcurve of
that period. Results are shown in Figure 1. Table 1 presents
the values of the reduced $\chi^{2}$ for fitting different models to the
modulation fraction and the peak flux in X-rays. It compares the
results of fitting a horizontal line, a linear fit, and two sinusoidal
functions. One of the latter has the same period and phase of
the radio modulation (from Gregory 2002, labeled as ¡°Radio¡±
in Table 1, dotted line in Figure 1). The other sine function has
the same period as in radio but allowing for a phase shift from it
(a solid line in Figure 1, labeled as ¡°Shifted¡± in Table 1). It is
clear that there is variability in the data and the sinusoidal description with a phase
shift is better than the linear one. The phase shift derived by fitting the modulated
fraction is 281.8$\pm$44.6 days, corresponding in phase
to $\sim$0.2 of the 1667$\pm$8 day super-orbital period. The phase
shift derived by fitting the maximum flux is 300.1$\pm$39.1 days,
which are compatible with the former.

\section{Conclusion}

We have found evidence of super--orbital modulation in the X-ray emission from \lsi.
We show that there is a $\sim0.2$ phase shift between the radio and the X-ray super-orbital modulation. Torres et al. 2012 has proposed that \lsi\ could be subject to
a flip-flop behavior. The super¨Corbital modulation is possibly due to
the cyclic change of the circumstellar disk (Li et al. 2012; see also
Papitto et al. 2012). In this context, multi--wavelength super--orbital modulation is expected and confirmed in radio, optical, X-ray and hinted in TeV (Li et al. 2012).

\end{document}